\documentclass[10pt,letterpaper]{article}

\usepackage{spconf}
\usepackage{latexsym}
\usepackage{amsmath,amssymb}
\usepackage[T1]{fontenc}
\usepackage[utf8]{inputenc}
\usepackage{microtype}
\usepackage{graphicx}
\usepackage{booktabs}
\usepackage{tabularx}
\usepackage[authoryear,round]{natbib}
\usepackage[hidelinks]{hyperref}

\title{From Spatial Perception to Speech Recovery: Binaural Semantic Adaptation for Spatially Grounded Multi-Talker Transcription}
\name{\normalfont Wenxuan He$^{1,2}$ \qquad Shan Liang$^{1,*}$}
\address{\normalsize $^1$\,Xi'an Jiaotong-Liverpool University\\
\normalsize $^2$\,University of Liverpool\\[4pt]
\small \href{mailto:Wenxuan.He23@xjtlu.edu.cn}{Wenxuan.He23@xjtlu.edu.cn}\quad
\href{mailto:Wenxuan.He@liverpool.ac.uk}{Wenxuan.He@liverpool.ac.uk}\\[1pt]
\small $^*$Corresponding author: \href{mailto:Shan.Liang@xjtlu.edu.cn}{Shan.Liang@xjtlu.edu.cn}}

\begin{document}
\maketitle

\begin{abstract}
Spatial audio large language models increasingly combine general audio understanding with spatial perception. Yet overlapping conversations require more than locating speakers. A model must also recover what each person says and match the words to their positions. We examine this challenge in MiDashengLM-Spatial, a binaural model designed for both spatial reasoning and general audio understanding. To test these abilities, we introduce BiCP-Bench, a benchmark of overlapping two-speaker binaural recordings that contrasts candidate-based recognition, position-guided transcription, and joint transcription with spatial attribution. The model can identify a target utterance among candidates but struggles to transcribe it without them. Its semantic encoder uses only one ear. Changing that ear shifts which side's speech is recovered more accurately, despite unchanged binaural input to the spatial branch. We therefore propose Binaural Semantic Adaptation (BiSA), which combines shared encoding of both ears with cross-ear differential interactions. With the backbone frozen and 0.027\% trainable parameters, BiSA raises correctly attributed joint transcription success from 0\% to 57.81\% for simultaneous onsets. For staggered onsets, later-speaker word error rate falls from 109.47\% to 29.58\%. On full-length Clarity CEC1 mixtures, word error rate falls from 59.33\% to 39.60\%. Spatial reasoning on MMAU-Pro improves by 9.54 percentage points without updating the spatial encoder, and general audio performance remains largely intact.
\end{abstract}

\begin{keywords}
Spatial audio understanding, cocktail party problem, overlapping speech, multi-talker transcription, spatial attribution.
\end{keywords}

\section{Introduction}
\label{sec:introduction}
\begin{figure*}[t]
\centering
\includegraphics[width=0.90\textwidth]{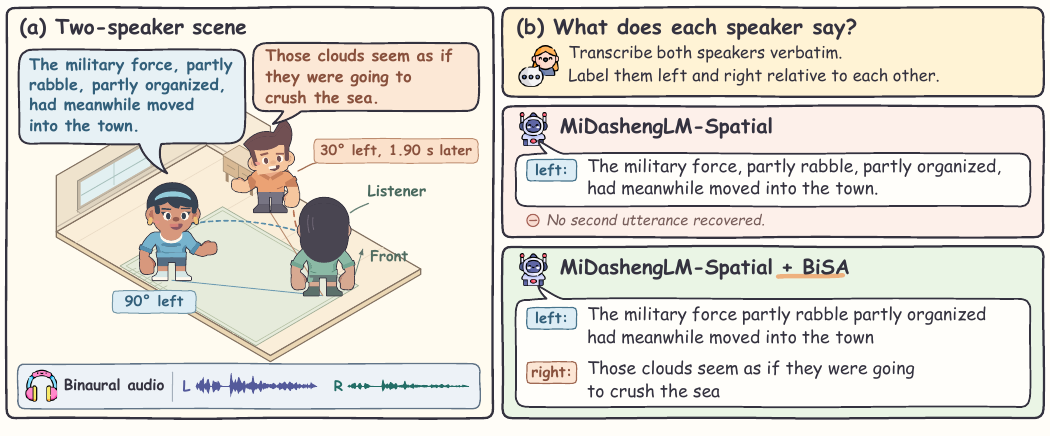}
\caption{Joint transcription in a same-side BiCP-Async mixture. Both speakers are to the listener's left. The speaker at \(30^\circ\) starts 1.90~s after the speaker at \(90^\circ\). MiDashengLM-Spatial recovers only the earlier speaker's utterance. BiSA recovers both utterances, assigning relative \texttt{left} to the \(90^\circ\) speaker and \texttt{right} to the \(30^\circ\) speaker.}
\label{fig:bicp-example}
\end{figure*}

Spatial hearing enables listeners to locate sound sources and understand their content in complex acoustic environments. In multi-talker conversations, spatial cues help listeners segregate overlapping speech and associate individual utterances with their speakers. In recent years, audio large language models (LLMs) have made substantial progress in speech recognition, sound event recognition, and audio question answering. BAT \citep{zheng2024bat} and OWL \citep{biswas2026owl} have further extended the capabilities of audio LLMs to binaural spatial perception, enabling sound source localization and reasoning about spatial relations. More recently, MiDashengLM-Spatial \citep{hu2026midasheng} has integrated spatial perception and general audio understanding within a unified architecture, advancing the development of spatial audio LLMs.

Advances in spatial perception open new opportunities for audio LLMs to understand complex multi-talker conversations. Previous studies have shown that spatial information can guide the recognition of speech from a specified direction \citep{tang2024spatial,dementyev2026phasecoder}. A further challenge in overlapping speech is to recover each speaker's complete utterance and associate it with the correct spatial position. We refer to this task as \emph{spatially grounded multi-talker transcription}. In the same-side mixture shown in Figure~\ref{fig:bicp-example}, the original model recovers the earlier speaker's utterance but omits the later speaker's speech. This raises a central question: can existing audio LLMs translate their spatial perception capabilities into spatially grounded speech understanding?

To investigate this question, we introduce the Binaural Cocktail Party Benchmark (BiCP-Bench). The benchmark covers two-speaker mixtures with simultaneous or staggered speech onsets. It evaluates candidate-based utterance selection, position-conditioned target-speaker transcription, and joint transcription with spatial attribution. We study the recently introduced MiDashengLM-Spatial, an advanced binaural audio LLM with strong spatial perception and general audio understanding capabilities. Our experiments show that the model can select a target utterance from candidates given the speaker's position but struggles to produce its complete transcript through free-form transcription. When both speakers start speaking simultaneously, the original model fails to produce a single successful joint transcription with correct spatial attribution. These findings reveal a gap between spatial perception, candidate-assisted recognition, and open-ended speech recovery.

We further examine how semantic encoding contributes to this capability gap. Although MiDashengLM-Spatial receives binaural audio, its semantic encoder processes only one ear signal. We switch the semantic input from the left ear to the right while keeping the spatial branch's acoustic input unchanged. The transcription advantage shifts from the speaker on the left to the speaker on the right. We further compare complementary binaural representations with repeated single-ear representations at the same input length. Providing representations from both ears improves the recovery of both speakers' utterances. These findings show that the two ears provide complementary speech information about the same mixture, which single-ear semantic encoding cannot fully exploit. Sound from each speaker reaches the two ears differently depending on the speaker's position, providing an acoustic basis for this complementarity. This motivates us to exploit cross-ear acoustic differences during semantic encoding.

Building on these findings, we propose Binaural Semantic Adaptation (BiSA). The core idea is to use complementary observations of the same acoustic scene and incorporate binaural acoustic differences into the construction of semantic representations. BiSA uses a shared semantic encoder to extract representations from both ears and introduces cross-ear differential interactions to enhance speech content representations during encoding. Training first uses position-conditioned target-speaker and joint transcription supervision to learn complete utterance recovery. A second stage combines transcription and spatial selection tasks to calibrate the model's spatial responses. BiSA improves spatially grounded multi-talker transcription by training a small set of added parameters while keeping the original backbone frozen.

BiSA adds trainable parameters amounting to only 0.027\% of the original model's parameter count. On BiCP-Bench, joint transcription success with correct spatial attribution rises from 0\% to 57.81\% when both speakers start simultaneously. Under staggered overlap, joint transcription success with correct spatial attribution rises from 0\% to 53.65\%. WER for the later-starting speaker decreases from 109.47\% to 29.58\%. BiSA also generalizes to the external Clarity CEC1 dataset, reducing target-speaker WER on untrimmed mixtures from 59.33\% to 39.60\%. It largely preserves the original model's general audio understanding and maintains near-baseline spatial reasoning performance on the STAR-Bench spatial-reasoning subset. With the spatial encoder frozen, BiSA improves spatial reasoning accuracy on MMAU-Pro by 9.54 percentage points. These findings suggest that the information preserved during semantic encoding can also shape spatial understanding. By exploiting complementary binaural speech cues, BiSA improves utterance recovery, spatial attribution, and spatial reasoning while keeping the spatial encoder frozen.

Our main contributions are as follows:

\begin{enumerate}
\item \textbf{Evaluation of spatially grounded speech understanding.} We introduce BiCP-Bench to evaluate candidate-based utterance selection, position-conditioned target-speaker transcription, and joint transcription with spatial attribution under overlapping speech. Our evaluation reveals a gap between spatial perception and speech recovery.
\item \textbf{Analysis of reference-ear dependence.} We show that single-ear semantic encoding biases speech recovery toward speakers at particular positions, with the transcription advantage shifting when we switch the reference ear. Controlled experiments further demonstrate the value of complementary binaural representations for recovering both speakers' utterances.
\item \textbf{Parameter-efficient binaural semantic adaptation.} We propose BiSA, which incorporates cross-ear differential interactions into semantic encoding. With the original backbone frozen, BiSA substantially improves spatially grounded multi-talker transcription and enhances spatial reasoning using only a small set of trainable parameters. These improvements further reveal that semantic representation construction is a key stage of spatial–semantic interaction.
\end{enumerate}

\section{Related Work}
\label{sec:related-work}

\subsection{Spatial Audio Large Language Models}
\label{sec:related-spatial-llms}

Recent audio large language models have expanded from general sound recognition and question answering to spatial audio understanding. BAT \citep{zheng2024bat} combines a binaural spatial encoder with a large language model to support sound event recognition, source localization, and spatial relation reasoning. OWL \citep{biswas2026owl} introduces geometric supervision and multi-step spatial reasoning to improve the understanding of source direction, distance, and relative spatial relations. DSpAST \citep{wilkinghoff2026dspast} improves spatial reasoning through disentangled spatial acoustic representations. PhaseCoder \citep{dementyev2026phasecoder} uses multichannel audio and microphone coordinates to construct array-independent spatial representations for spatial question answering and target-speaker transcription. More recently, MiDashengLM-Spatial \citep{hu2026midasheng} combines semantic and spatial encoding in a unified model for general audio understanding and binaural spatial perception. These studies have advanced spatial audio models from source perception toward more complex spatial understanding, providing a foundation for spatially grounded speech recovery in overlapping speech.

\subsection{Spatially Conditioned and Multi-Talker Speech Recognition}
\label{sec:related-multi-talker}

Overlapping speech recognition requires recovering the content of multiple speakers and assigning each utterance to the corresponding speaker. Serialized output training \citep{kanda2020sot} organizes transcripts from multiple speakers into a single output sequence for end-to-end multi-talker recognition. \citet{meng2025mtllm} extend instruction-driven language models to joint multi-talker transcription and target transcription conditioned on speaker attributes. \citet{tang2024spatial} use spatial cues for far-field speech recognition and target speech extraction from specified directions. \citet{xie2025directivity} introduce directional markers and multichannel training to enable speech language models to jointly handle transcription and source direction information. \citet{lin2026directional} investigate cascaded systems incorporating directional speech separation alongside end-to-end approaches based on serialized outputs. More recently, \citet{polok2026dixtral} condition a semantic encoder on speaker activity to support target transcription and content understanding in multi-talker speech. These studies demonstrate the value of spatial conditions, speaker cues, and output organization for overlapping speech processing. Building on this research, we examine binaural semantic representations in pretrained spatial audio models. We investigate how complementary cross-ear information during semantic encoding can improve multi-talker speech recovery and spatial attribution while retaining the original spatial encoding pathway.

\subsection{Spatial Audio and Multi-Talker Understanding Benchmarks}
\label{sec:related-benchmarks}

Spatial audio evaluation has expanded from sound source localization to spatial relations and language understanding. SpatialSoundQA \citep{zheng2024bat} introduces questions covering sound event perception, direction estimation, distance estimation, and spatial relation reasoning. BiDepth \citep{biswas2026owl} further incorporates geometric information and supervision for multi-step spatial reasoning. For broader audio understanding, MMAU-Pro \citep{kumar2026mmau} covers speech, environmental sounds, music, and spatial reasoning. More recently, MSU-Bench \citep{sun2026msu} introduces speaker reference, content understanding, and cross-speaker reasoning tasks for multi-party conversations, using multiple-choice questions to evaluate speaker-centric understanding. These benchmarks provide complementary perspectives on spatial audio and multi-talker understanding. Our BiCP-Bench further distinguishes candidate-based utterance recognition, open-ended recovery of complete utterances, and spatial attribution under spatial conditions. Through simultaneous and staggered overlap scenarios, it evaluates whether models can reliably associate each speaker's speech content with the corresponding spatial position.

\section{BiCP-Bench: Evaluating Spatially Grounded Multi-Talker Transcription}
\label{sec:benchmark}

Performance on sound source localization and spatial question answering does not establish whether a model can fully transcribe overlapping speakers. This requires recovering each speaker's words and associating them with the correct position within the same mixture. The Binaural Cocktail Party Benchmark (BiCP-Bench) measures this gap by contrasting candidate-based recognition with generated transcripts and assessing their spatial attribution. Its two-speaker binaural mixtures cover simultaneous and staggered onsets, with speakers on opposite sides, on the same side, or directly ahead and to one side.

\subsection{Task Formulation}
\label{sec:bicp-tasks}

The three tasks vary the linguistic support and output requirements. They progress from selecting a supplied utterance to generating one speaker's transcript, then jointly transcribing both speakers with spatial attribution.

\noindent\textbf{Candidate-based Content Recognition.} The first task probes speech content recognition given the target speaker's position. Along with a binaural mixture and the target position, the model receives four candidate sentences and selects the one spoken by the target speaker. Both speakers' reference utterances are included, together with two distractors of similar length drawn from other utterance pairs. Including the other speaker's utterance makes recognizing speech content alone insufficient: the model must also identify its spatial origin. We evaluate candidate-based recognition on BiCP-Sync.

\noindent\textbf{Position-conditioned Target Transcription.} Recovering speech without candidate sentences poses a further challenge. The model receives the complete binaural mixture and a natural language instruction specifying the target position. Its output is the target speaker's complete utterance. We query both speakers separately to examine whether the model can recover the corresponding utterance when the spatial instruction changes.

\noindent\textbf{Joint Transcription with Spatial Attribution.} The model must now recover both speakers' utterances and attribute them to their respective positions in a single response. Given the binaural mixture, it transcribes both utterances in full and labels the more leftward and more rightward speakers. Unlike separate target queries, joint transcription requires the model to organize the two utterances and establish their spatial associations within one generation.

\noindent\textbf{Spatial Position Descriptions.} We use absolute directions to describe source locations relative to the listener and relative positions to distinguish the two speakers from each other. The frontal direction is defined as 0\textdegree{}, with positive azimuths to the left and negative azimuths to the right. Of the two speakers, the one with the larger azimuth is designated as the more leftward speaker and the other as the more rightward speaker. For instance, both speakers at +90\textdegree{} and +30\textdegree{} lie to the listener's left, but the former is more leftward than the latter.

Candidate-based recognition uses absolute directions. In BiCP-Sync, target transcription uses absolute directions for speakers on opposite sides and relative positions for speakers on the same side. Relative positions are used throughout target transcription in BiCP-Async and joint transcription in both subsets. They allow us to assess speech recovery even when a coarse directional label cannot distinguish the speakers.

\subsection{Benchmark Construction}
\label{sec:bicp-construction}

We construct binaural mixtures with variations in speaker positions, speech levels, and onset times to assess spatially grounded speech understanding across different acoustic conditions. BiCP-Bench comprises two subsets distinguished by the speakers' onset times. BiCP-Sync contains 1,152 simultaneous-onset mixtures, and BiCP-Async contains 384 staggered-onset mixtures.

\noindent\textbf{Speech Corpora and Binaural Rendering.} We draw clean utterances from LibriSpeech test-clean \citep{panayotov2015librispeech} and the English test split of FLEURS \citep{conneau2023fleurs}. The binaural responses come from ARI and the University of Surrey. ARI provides measured head-related impulse responses (HRIRs) from different listeners \citep{majdak2010localization}. Surrey provides binaural room impulse responses (BRIRs) measured using a dummy head \citep{francombe2017surrey}. Reverberant conditions use room responses simulated with pyroomacoustics \citep{scheibler2018pyroom} and ARI directional responses, or complete measured Surrey BRIRs. Table~\ref{tab:bicp-composition} summarizes the three combinations of speech corpora and binaural responses, covering different speech content, listener anatomies, and acoustic environments.

Each clean utterance is convolved with the left- and right-ear impulse responses corresponding to its source azimuth. We scale and mix the resulting signals according to the prescribed speech levels and onset times. The signal received at ear \(e\) is given by:

\begin{equation}
\label{eq:binaural-mixture}
x_e(t)=g\sum_{i=1}^{2}a_i(s_i*h_{i,e})(t-\tau_i),\quad e\in\{L,R\},
\end{equation}

In Eq.~\eqref{eq:binaural-mixture}, \(s_i\) denotes the clean utterance of speaker \(i\), and \(h_{i,e}\) denotes the acoustic impulse response from that speaker to ear \(e\). The coefficients \(a_i\) and \(\tau_i\) control the speech level and onset time, respectively. The gain \(g\) is applied equally to both channels after mixing. All mixtures are stored as two-channel audio at 16 kHz, preserving the acoustic differences between the ears.

\begin{table}[tb]
\centering
\caption{Composition of BiCP-Bench.}
\label{tab:bicp-composition}
\par\vspace{4pt}
\small
\setlength{\tabcolsep}{4pt}
\begin{tabular}{@{}llrr@{}}
\toprule
Speech corpus & \shortstack[l]{Binaural\\responses} & BiCP-Sync & BiCP-Async \\
\midrule
\addlinespace[2pt]
LibriSpeech & ARI & 384 & 128 \\
FLEURS & ARI & 384 & 128 \\
FLEURS & Surrey & 384 & 128 \\
\multicolumn{2}{@{}l}{Distinct clean utterance pairs} & 128 & 188 \\
\midrule
\addlinespace[2pt]
\multicolumn{2}{@{}l}{Total mixtures} & 1,152 & 384 \\
\bottomrule
\end{tabular}
\end{table}

In BiCP-Sync, the FLEURS groups rendered with ARI and Surrey share the same utterance pairs, source azimuths, nominal gains, and evaluation questions. This pairing enables comparisons across different binaural responses. BiCP-Async distributes acoustic conditions across different source pairs to cover a wider range of utterances and speaking orders.

\noindent\textbf{BiCP-Sync.} Aligning the speakers' speech onsets reduces the influence of speaking order. High overlap tests whether the model can recover both complete utterances while the speakers are active together. We trim leading and trailing silence and select utterance pairs of similar duration with substantial overlap in their active speech regions. Both trimmed utterances start at the same sample. We measure overlap as the smaller of the two proportions of active speech frames shared with the other utterance. The minimum overlap rates among the selected LibriSpeech and FLEURS pairs are 90.78\% and 77.09\%, respectively.

Within each rendering group, six acoustic variants per utterance pair vary spatial separation, speech levels, and reverberation to examine speech recovery under different conditions. The basic condition places the speakers at \ensuremath{\pm}30\textdegree{} with equal gains after active-speech level normalization. In two further conditions, one speaker's clean signal is attenuated by 4 dB or 8 dB to examine recovery of weaker speech. To test relative spatial distinctions within one side, we place both speakers at +30\textdegree{} and +90\textdegree{} or \ensuremath{-}30\textdegree{} and \ensuremath{-}90\textdegree{}. Distinguishing these speakers requires finer spatial information than coarse left–right classification. The remaining conditions are a reverberant mixture and a channel-swapped counterpart of the 8 dB attenuation condition.

Each acoustic variant contains 192 mixtures, yielding 1,152 simultaneous-onset mixtures in total. The speakers are on opposite sides of the listener in 960 mixtures. In the other 192, both are on the left or both are on the right.

\noindent\textbf{BiCP-Async.} By introducing staggered speech onsets, BiCP-Async examines whether the model can recover speech from a specified position after another speaker has already begun speaking.

At each ear, the active speech onsets differ by 1–2 seconds. At least 80\% of the later speaker's active speech overlaps with the earlier speaker's active speech. The earlier speaker also finishes 0.5–1 second after the later speaker, measured from the end of active speech at each ear. Most of the later utterance is therefore embedded within the earlier speaker's ongoing speech. We retain both utterances in full, including the earlier speaker's speech before and after the overlap.

The spatial configurations include speakers on opposite sides, one directly ahead and the other to one side, and both on the same side. Opposite-side mixtures test left–right separation, whereas the other configurations require distinguishing positions within one side or between the front and a side. Varying the strength of the later speaker's speech tests recovery when temporal and level differences occur together. During overlapping active speech, the later-to-earlier speaker energy ratio, aggregated over both ears, is set to 0, \ensuremath{-}4, or \ensuremath{-}8 dB.

Each of the 192 original staggered-onset scenes has a channel-swapped counterpart, giving 384 binaural mixtures. Unlike BiCP-Sync, which expands each utterance pair across multiple acoustic variants, BiCP-Async uses more distinct utterance pairs to cover a wider variety of speech content.

\noindent\textbf{Spatial Mirroring.} Channel swapping provides paired scenes for examining whether the model responds to changes in spatial position. Swapping the left and right channels preserves both utterances, their speaking order, and their timing. It reverses the represented source azimuths and corresponding spatial labels. In BiCP-Sync, we apply this operation to the 8 dB attenuation condition; in BiCP-Async, we apply it to every original scene. These pairs allow us to examine whether changes in spatial information lead to corresponding changes in the model's attribution of speech content.

\noindent\textbf{Data Isolation.} We separate the adaptation training data from the BiCP-Bench test data to evaluate generalization to unseen speech content and listener anatomies. There is no overlap in source recording identifiers, normalized reference utterances, or verifiable speaker identifiers between the training and test sets. Speaker-level disjointness cannot be established for FLEURS because persistent speaker identifiers are unavailable in its public manifest. The ARI listeners used for training, validation, and testing are mutually disjoint. BiCP-Sync and BiCP-Async share some source speech material, so benchmark size is reported in terms of the resulting binaural mixtures.

\subsection{Evaluation Protocol}
\label{sec:bicp-evaluation}

Candidate-based content recognition is measured by the proportion of correctly answered queries. This accuracy measures whether the model selects the utterance associated with the requested position when candidate content is available.

\noindent\textbf{Target Transcription.} We measure the discrepancy between the generated text and the target speaker's reference utterance using word error rate (WER):

\begin{equation}
\label{eq:wer}
\mathrm{WER}=\frac{S+D+I}{N},
\end{equation}

In Eq.~\eqref{eq:wer}, \(S\), \(D\), and \(I\) denote the numbers of word substitutions, deletions, and insertions, respectively, and \(N\) is the number of reference words. Corpus-level WER is the total number of word errors divided by the total number of reference words. WER values are reported as percentages.

A target transcription is considered successful when its WER does not exceed 20\%. We report the proportion of queries meeting this criterion. For mixtures with two separate target queries, we also report the proportion of mixtures for which both transcriptions succeed. On BiCP-Async, results for the earlier and later speakers are reported separately to examine speech recovery under different speaking orders.

\noindent\textbf{Joint Transcription.} Recovering both utterances and attributing them to the correct positions are distinct aspects of joint transcription. We measure them separately to distinguish incomplete speech recovery from spatial attribution errors.

We first extract utterances and explicit position labels from the response. Content recovery asks whether both speakers' utterances are present, regardless of their spatial labels. We find the best assignment of distinct output segments to the two reference utterances. A mixture counts as successful if an assignment achieves WER of at most 20\% for both utterances and leaves no unmatched output words.

For joint transcription with correct spatial attribution, we compare the text assigned to each explicit left or right label against the corresponding reference utterance. Multiple segments under the same label are concatenated in their generated order. A mixture succeeds when the WER for each position does not exceed 20\%. This metric tests whether the recovered utterances are associated with their correct relative positions. The two success rates are computed and reported separately because their text matching rules differ.

We also report overall WER for joint transcription. The assignment between the reference utterances and distinct output segments minimizes the total number of word errors, counting words in unmatched segments as insertions. Missing utterances are matched to empty segments and counted as deletions. Corpus-level WER is computed over all reference words.

This protocol provides controlled comparisons for the following analysis of where spatial perception falls short of recovering and attributing overlapping speech.

\section{Diagnosing the Speech Recovery Gap}
\label{sec:recovery-gap}

A model may locate sound sources without recovering the corresponding speech content. This gap is particularly important in overlapping speech. Even with spatial information available, the model may fail to recover complete utterances from speakers at different locations. We therefore examine whether the model's ability to access speech content limits spatially grounded transcription.

\subsection{Recognition Does Not Imply Speech Recovery}
\label{sec:recognition-recovery}

MiDashengLM-Spatial shows markedly different performance on candidate-assisted recognition and free-form transcription. BiCP-Sync contains 1,152 binaural recordings of two speakers with simultaneous onsets and overlapping speech. Both tasks query each speaker separately in every recording, yielding 2,304 individually scored responses per task. Given a target position, the model selects the corresponding utterance from four candidates. It achieves 69.27\% accuracy (1,596/2,304), above the 25\% chance level. Without candidate sentences, only 26.91\% (620/2,304) of position-conditioned transcriptions meet the success criterion of a word error rate (WER) no greater than 20\%. The model can identify target utterances from candidates, but recovering their complete content without candidates remains difficult.

Recovering both speakers' utterances from the same mixture proves particularly difficult. For each of the 1,152 recordings, we query the two speakers separately, requesting one speaker's transcript at a time. None of these recordings yields two successful transcriptions. In joint transcription, the model must provide both utterances in one response and assign each to the correct relative left or right position. The baseline succeeds on 0 of 1,152 recordings (0\%). To examine whether attribution errors alone explain these failures, we also evaluate the same joint responses for content recovery without using spatial labels. Generated utterances are matched to the two reference transcripts using the best assignment, regardless of their predicted left or right labels. Under this measure, successful recovery of both utterances also remains at 0/1,152.

These results point to a limitation in speech content recovery itself. The model struggles to recover both utterances from overlapping speech. Candidate sentences help it identify target content, but complete free-form transcription remains unreliable. This recovery gap directs our attention to how the model constructs semantic representations.

\subsection{Semantic Reference-Ear Dependence}
\label{sec:reference-ear}

MiDashengLM-Spatial receives binaural audio, but its semantic encoder and spatial branch use different acoustic inputs. The spatial branch processes both ears, while the semantic encoder uses only the fixed left-ear signal. The two ears record acoustically different versions of the same overlapping speech. We investigate the relationship between semantic input and transcription by switching the semantic reference ear from left to right. The original binaural recording and the spatial branch's input remain unchanged.

We compare 200 recordings containing two overlapping speakers, keeping the model weights and other inference settings fixed. Both conditions use joint transcription, requiring the model to produce the two speakers' utterances in one response. We assess content recovery for left- and right-side utterances separately, using a WER of at most 20\% as the success criterion. Scoring ignores spatial labels and extra output, checking whether each speaker's content meets the transcription criterion.

\begin{table}[tb]
\centering
\caption{Speech content recovery under semantic reference-ear switching.}
\label{tab:reference-ear-recovery}
\par\vspace{4pt}
\small
\setlength{\tabcolsep}{4pt}
\begin{tabular}{@{}lcc@{}}
\toprule
\shortstack[l]{Semantic\\reference ear} & \shortstack{Left-side utterances\\recovered} & \shortstack{Right-side utterances\\recovered} \\
\midrule
\addlinespace[2pt]
Left ear & 162/200 (81\%) & 4/200 (2\%) \\
Right ear & 4/200 (2\%) & 172/200 (86\%) \\
\bottomrule
\end{tabular}
\end{table}

Table~\ref{tab:reference-ear-recovery} shows that switching the semantic reference ear produces a clear reversal in content recovery performance. With left-ear input, 162 of the 200 left-side utterances meet the criterion, compared with four from the right side. Switching to right-ear input yields 172 successful right-side utterances and four successful left-side utterances. Which speaker's speech is recovered more accurately depends strongly on whether the semantic encoder uses the left or right ear.

The experiment reveals a strong dependence of speech content recovery on the semantic reference ear. The spatial branch receives the same binaural input in both conditions. Changing the semantic reference ear alone shifts which side's utterance is recovered more reliably. This provides a more specific clue to the recovery gap observed in Section~\ref{sec:recognition-recovery}. The existing semantic pathway does not fully exploit the different observations of the same mixture available at the two ears. Whether these observations contain complementary information remains to be tested.

\subsection{Complementary Semantic Information Across Ears}
\label{sec:ear-complementarity}

To test whether the two ears provide complementary information, we compare binaural semantic representations with repeated single-ear representations.

We keep the model weights, mixed recordings, spatial features, and task instructions fixed. We compare four semantic input combinations: left–right, right–left, left–left, and right–right. The first two combine representations from different ears. The remaining two repeat representations from a single ear, and all four inputs have the same sequence length. We evaluate joint transcription on 200 recordings with overlapping speech from two speakers, examining whether the responses contain both speakers' complete utterances. Generated utterances are matched to the two reference transcripts using the best assignment. Recovery succeeds when both speakers achieve a WER of at most 20\%, regardless of spatial labels.

\begin{table}[tb]
\centering
\caption{Two-speaker content recovery with binaural and repeated single-ear representations.}
\label{tab:ear-complementarity}
\par\vspace{4pt}
\small
\setlength{\tabcolsep}{5pt}
\begin{tabular}{@{}lc@{}}
\toprule
Input representations & \shortstack{Successful two-speaker\\content recovery} \\
\midrule
\addlinespace[2pt]
Left + Right & 66/200 \\
Right + Left & 65/200 \\
Left + Left & 2/200 \\
Right + Right & 3/200 \\
\bottomrule
\end{tabular}
\end{table}

Table~\ref{tab:ear-complementarity} shows that both binaural combinations recover the two speakers' utterances more often than repeated single-ear representations. This indicates that the two ears provide complementary information useful for recovering individual utterances from overlapping speech.

Together with the reference-ear experiment, these results show that the ear chosen for semantic encoding affects which speaker's content is recovered. Using both ears improves recovery of the two utterances. These findings motivate changes to the current single-ear semantic encoding pathway. How can we use the complementary information from both ears when constructing semantic representations to improve the recovery of individual utterances in overlapping speech? This question motivates the method developed in the next section.

\section{Binaural Semantic Adaptation (BiSA)}
\label{sec:binaural-adaptation}

\begin{figure*}[t]
\centering
\includegraphics[width=0.90\textwidth]{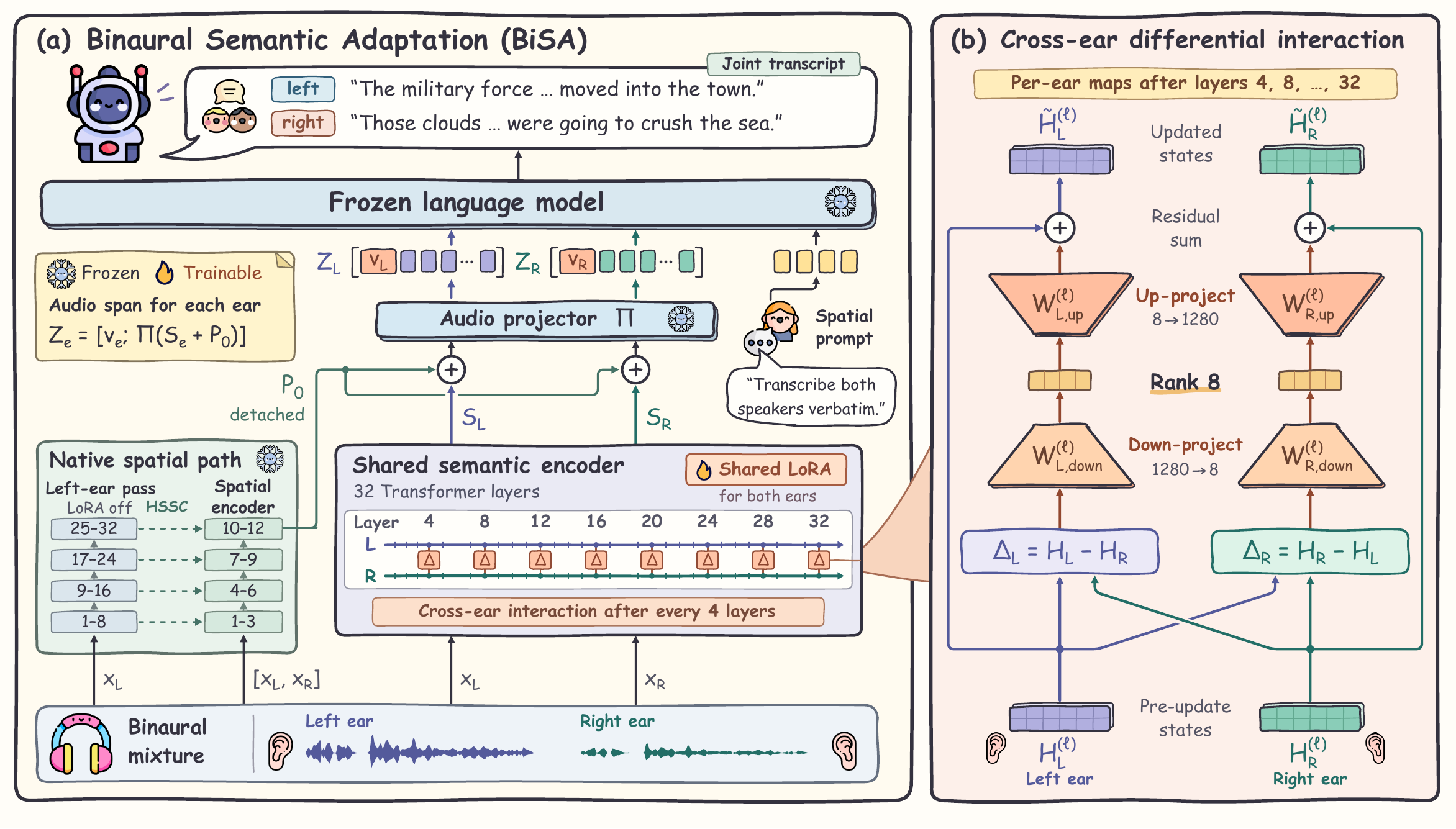}
\caption{BiSA architecture and cross-ear differential interaction. (a) A shared semantic encoder processes both ears with shared LoRA and cross-ear residual modules. The detached native spatial contribution \(P_0\) is added to both semantic sequences before projection. Each projected audio span receives an ear-identity vector and enters the frozen language model. All pretrained weights remain frozen. (b) Independent rank-8 maps transform each ear's difference from the other ear into a residual update after layers 4, 8, \ldots, 32. Both updates use the same pre-update states.}
\label{fig:bisa-architecture}
\end{figure*}

The analysis in Section~\ref{sec:recovery-gap} shows that the choice of semantic reference ear strongly affects the model's ability to recover speech from different speakers. Complementary information from the two ears also helps recover both utterances in overlapping speech. Based on these findings, we propose Binaural Semantic Adaptation (BiSA) to improve spatially grounded multi-talker transcription by changing how semantic representations are constructed. BiSA extends the original single-ear semantic encoding to shared-weight binaural encoding and introduces cross-ear differential interaction during encoding, allowing acoustic observations from both ears to contribute jointly to semantic representation construction. BiSA leaves all pretrained weights unchanged, including those of the audio projector and language model. The adaptation is implemented through added LoRA parameters, cross-ear differential modules, and two identity vectors. These parameters can be stored and loaded separately from the pretrained backbone. Figure~\ref{fig:bisa-architecture} summarizes the architecture and the cross-ear residual update.

\subsection{Binaural Semantic Encoding}
\label{sec:binaural-semantic-encoding}

The semantic encoder in MiDashengLM-Spatial uses a fixed left-ear signal, while the spatial branch independently processes the full binaural audio. To exploit the different acoustic observations captured by the two ears, BiSA feeds the left- and right-ear signals separately into a shared semantic encoder. Both ears use the same pretrained encoder weights to extract their respective semantic representations in a common feature space.

Binaural input provides two sets of semantic representations, but the pretrained encoder has not been adapted to recovering overlapping speech from binaural recordings. We therefore freeze its pretrained weights and introduce shared low-rank adaptation (LoRA) parameters \citep{hu2022lora}. These parameters allow the encoder to adjust its feature transformations under transcription supervision while retaining its pretrained representation capabilities. The two ears use the same LoRA parameters. The original spatial encoding branch remains frozen and continues to extract spatial information independently from the binaural audio.

\subsection{Cross-Ear Differential Interaction}
\label{sec:cross-ear-interaction}

With shared binaural encoding, the left- and right-ear representations are still computed along separate paths, without explicit modeling of their relationship during encoding. To model this relationship explicitly during semantic encoding, BiSA introduces cross-ear differential interaction. The two ears capture different acoustic observations of the same scene, and the differences between their representations provide a basis for modeling this relationship. We use hidden-state differences to construct interaction signals, allowing each ear's semantic representation to be updated using information from the other ear.

Let \(H_L^{(\ell)},H_R^{(\ell)}\in\mathbb{R}^{n_\ell\times d}\) denote the left- and right-ear hidden states at layer \(\ell\), respectively. Here, \(n_\ell\) is the number of aligned audio tokens at layer \(\ell\). The cross-ear differential module computes the differences between the two representations at corresponding token positions:

\begin{equation}
\label{eq:cross-ear-difference}
\Delta_L^{(\ell)}=H_L^{(\ell)}-H_R^{(\ell)},\qquad
\Delta_R^{(\ell)}=H_R^{(\ell)}-H_L^{(\ell)}.
\end{equation}

For \(e\in\{L,R\}\), the residual update using Eq.~\eqref{eq:cross-ear-difference} is

\begin{equation}
\label{eq:cross-ear-update}
\widetilde H_e^{(\ell)}
=
H_e^{(\ell)}
+
\Delta_e^{(\ell)}
\bigl(W_{e,\mathrm{down}}^{(\ell)}\bigr)^{\top}
\bigl(W_{e,\mathrm{up}}^{(\ell)}\bigr)^{\top}.
\end{equation}

In Eq.~\eqref{eq:cross-ear-update}, \(W_{e,\mathrm{down}}^{(\ell)}\in\mathbb{R}^{r\times d}\) and \(W_{e,\mathrm{up}}^{(\ell)}\in\mathbb{R}^{d\times r}\) are the down-projection and up-projection matrices, respectively, with hidden dimension \(d=1280\) and rank \(r=8\). The two ears and different layers use independent mapping parameters. Both residual updates use the same pre-update states. The updated states enter the next Transformer layer, or the encoder output processing at layer 32. The up-projection matrices are initialized to zero, so the cross-ear differential modules leave the original hidden states unchanged at the start of training.

The differential features from each ear pass through trainable low-rank mappings to produce residual updates for their respective hidden states. This allows each representation to incorporate cross-ear differences during encoding. We insert cross-ear differential modules after Transformer layers 4, 8, 12, 16, 20, 24, 28, and 32, enabling interaction at different levels of semantic representation.

Let \(S_e,P_0\in\mathbb{R}^{n_s\times d}\) denote the final semantic features and frozen native spatial features, respectively. The same \(P_0\) is added to both semantic sequences before the frozen audio projector \(\Pi\). The projected sequence for ear \(e\), prefixed by its identity vector, is
\begin{equation}
\label{eq:decoder-input}
Z_e=[v_e;\Pi(S_e+P_0)],\qquad e\in\{L,R\}.
\end{equation}
Here, \(n_s\) is the sequence length before projection, and \(n_a\) is the number of projected audio tokens per ear. The projector includes temporal subsampling. We have \(v_e\in\mathbb{R}^{1\times3584}\) and \(\Pi(S_e+P_0)\in\mathbb{R}^{n_a\times3584}\). Semicolons denote concatenation along the token axis. Equation~\eqref{eq:decoder-input} supplies the left and right audio spans to the frozen language model, retaining the native prompt delimiters. The native spatial features are computed with semantic LoRA disabled and detached from adaptation gradients.

\subsection{Position-Conditioned Supervised Learning}
\label{sec:position-supervision}

Spatially grounded multi-talker transcription requires the model to recover the utterances of different speakers, identify the target speaker from spatial information, and assign the recovered content to the correct position. We use two stages of supervised learning. The first stage focuses on position-conditioned transcription, training BiSA to generate the complete utterance of the speaker specified by a spatial instruction. To establish a consistent correspondence between positions and utterances, we adopt a spatial reference based on the speakers' relative azimuths, defining them as the more leftward and the more rightward speaker. For each two-speaker mixture, we construct separate transcription instructions for the two relative positions, using the corresponding speaker's complete utterance as the training target. We also introduce joint transcription supervision, requiring the model to recover both utterances in a single response and assign them according to the same relative left-right ordering.

The first-stage training examples include synchronous overlap, staggered overlap, and single-speaker speech to cover content recovery under different speaking conditions. We use an autoregressive text generation objective and compute the training loss against reference answers. Given binaural inputs \(x_L,x_R\), a task instruction \(q\), and a target answer sequence \(y=(y_1,\ldots,y_T)\), the loss is defined as

\begin{equation}
\label{eq:generation-loss}
\mathcal{L}_{\mathrm{gen}}
=-\frac{1}{T}\sum_{t=1}^{T}\log p_{\phi,\theta}(y_t\mid y_{<t},x_L,x_R,q),
\end{equation}

In Eq.~\eqref{eq:generation-loss}, \(\phi\) denotes the frozen pretrained parameters, \(\theta\) the adaptation parameters, and \(y_t\) the \(t\)-th answer token. The target sequence includes the end-of-sequence token. Prompt and audio-prefix tokens are excluded from the loss.

For a minibatch \(\mathcal{B}\), let \(\mathcal{Q}_b\) contain the supervised queries associated with record \(b\). The training objective averages token-normalized losses within each record and then across records:
\begin{equation}
\label{eq:record-loss}
\mathcal{L}_{\mathrm{batch}}=
\frac{1}{|\mathcal{B}|}\sum_{b\in\mathcal{B}}
\frac{1}{|\mathcal{Q}_b|}\sum_{q\in\mathcal{Q}_b}
\mathcal{L}_{\mathrm{gen}}^{(b,q)}.
\end{equation}
For paired target transcription, \(\mathcal{Q}_b\) contains one left query and one right query; other records contain one query. Equation~\eqref{eq:record-loss} therefore gives equal weight to the two target directions within each paired record.

In this stage, BiSA jointly optimizes the shared LoRA parameters in the semantic encoder, the cross-ear differential modules, and two learnable identity vectors. Because both ears use the same semantic encoder, we prepend a learnable vector to each semantic feature sequence to identify whether it comes from the left or right ear. During training, the pretrained weights of the semantic encoder, spatial encoding branch, and language model remain frozen. The generation loss is backpropagated through the frozen language model and projector to update the adaptation parameters; the native spatial features remain detached. BiSA has 2,300,928 trainable parameters in total, approximately 0.027\% of the base model.

Writing \(\theta=(\theta_{\mathrm{LoRA}},\theta_{\mathrm{diff}},v_L,v_R)\), Stage 1 updates all four components. After joint adaptation, we perform lightweight calibration of the two identity vectors. The second stage retains position-conditioned transcription, joint transcription, and noisy single-speaker transcription. It also introduces four-way single-speaker direction classification and spatially guided candidate-utterance selection, providing additional supervision for spatial recognition and the association between position and speech content. The two-speaker tasks continue to use the relative left-right reference from the first stage, giving position-conditioned transcription, joint transcription, and candidate-utterance selection a consistent correspondence between positions and utterances. In this stage, \(\theta_{\mathrm{LoRA}}\) and \(\theta_{\mathrm{diff}}\) are fixed, and only the two identity vectors, comprising 7,168 parameters, are updated. With the learned semantic encoding and cross-ear interaction parameters held fixed, this calibration adjusts how the language model reads semantic features from the two ears to better select and organize transcription output according to spatial instructions.

\section{Experimental Results and Analysis}
\label{sec:experiments}

Binaural recordings contain complementary cues about speech content. Whether cross-ear interactions during semantic encoding can exploit these cues to recover complete utterances from different speakers in overlapping speech warrants systematic investigation. This section examines BiSA on spatially grounded multi-talker transcription and further evaluates its cross-dataset applicability, effects on pretrained model capabilities, and the contribution of cross-ear interactions.

\subsection{Experimental Setup}

We use MiDashengLM-Spatial as the base model and compare the original model, untrained binaural input, and the two training stages of BiSA. The untrained binaural configuration encodes the left and right ears into separate semantic feature sequences without updating model parameters or using the trained cross-ear differential module. BiSA keeps the pretrained model parameters frozen and trains approximately 2.3 million adaptation parameters. The first stage uses 4,000 training records for binaural semantic adaptation. The second stage uses 480 calibration records and updates only the two ear-identity vectors. Models from both stages are selected using an independent development set.

Training binaural speech is constructed from LibriSpeech recordings and ARI head-related impulse responses. The main evaluation uses BiCP-Bench, comprising 1,152 simultaneous-overlap recordings (BiCP-Sync) and 384 staggered-overlap recordings (BiCP-Async). These recordings cover different relative speaker positions, speech materials, and acoustic response conditions. Training and test sets use different ear profiles, and overlap checks are performed on speech materials with verifiable identifiers. We also evaluate external speech recovery on Clarity CEC1 and assess general audio understanding and spatial capabilities using the STAR-Bench spatial-reasoning subset and MMAU-Pro. All models use the same task prompts, generation settings, and scoring protocols.

\subsection{Spatially Grounded Multi-Talker Transcription Performance}

To examine whether binaural semantic adaptation can turn complementary information from both ears into spatially grounded utterance recovery, we compare the original MiDashengLM-Spatial, untrained binaural input, and both BiSA stages on BiCP-Sync. The evaluation includes candidate-based utterance selection, position-conditioned transcription, and joint transcription of two speakers. Candidate-based selection asks the model to identify the utterance corresponding to a target speaker's position from four candidates. Position-conditioned transcription makes two independent requests for the same speech mixture, each specifying one speaker's position and requiring the complete utterance. Joint transcription requires the model to recover both complete utterances in a single response and assign their relative left-right positions. Table~\ref{tab:bicp-sync-results} presents the results.

\begin{table*}[t]
\centering
\caption{Spatially grounded transcription performance on BiCP-Sync}
\label{tab:bicp-sync-results}
\par\vspace{4pt}
\small
\setlength{\tabcolsep}{4pt}
\renewcommand{\arraystretch}{1.12}
\begin{tabularx}{\textwidth}{@{}l>{\centering\arraybackslash}X>{\centering\arraybackslash}X>{\centering\arraybackslash}X>{\centering\arraybackslash}X>{\centering\arraybackslash}X>{\centering\arraybackslash}X@{}}
\toprule
Model & Four-choice accuracy $\uparrow$ & Position-conditioned WER $\downarrow$ & Both-query success $\uparrow$ & Joint transcription WER $\downarrow$ & Both-speaker content success $\uparrow$ & Left-right attribution success $\uparrow$ \\
\midrule
\addlinespace[2pt]
Original model & 69.27 & 65.96 & 0.00 & 63.54 & 0.00 & 0.00 \\
Untrained binaural input & 58.33 & 66.15 & 1.39 & 54.23 & 5.12 & 5.03 \\
BiSA Stage 1 & 78.82 & \textbf{17.82} & 53.47 & 14.11 & 58.16 & 57.55 \\
BiSA Stage 2 & \textbf{80.51} & 18.12 & \textbf{53.56} & \textbf{14.02} & \textbf{58.85} & \textbf{57.81} \\
\bottomrule
\end{tabularx}
\par\smallskip
\begin{minipage}{\textwidth}
\footnotesize Note. All values are percentages. The evaluation contains 1,152 two-speaker mixtures, with 2,304 queries for the four-choice task. Both-query success requires the WER of each position-conditioned transcription for the same recording to be at most 20\%. For joint transcription, both-speaker content success uses optimal text matching, while left-right attribution success is scored using explicit position labels. The two metrics are computed independently.
\end{minipage}
\end{table*}

In candidate-based utterance selection, the original model achieves 69.27\% accuracy. However, in position-conditioned transcription without candidate utterances, it achieves no recordings with successful recovery of both speakers' complete utterances. After binaural semantic adaptation, BiSA Stage 2 increases candidate-selection accuracy to 80.51\%. Successful individual position-conditioned transcriptions increase from 620/2,304 to 1,675/2,304, and both-query success reaches 53.56\% (617/1,152). Overall WER decreases from 65.96\% to 18.12\%. Among 192 recordings with both speakers on the same side of the listener, BiSA successfully transcribes both speakers in 97 cases. The marked difference between candidate-based recognition and open-ended transcription reflects the distinct demands these tasks place on speech content recovery. Joint adaptation of binaural semantic representations substantially improves the recoverability of both speakers' utterances from the same mixture, enabling spatial instructions to guide complete utterance generation for a specified speaker.

Joint transcription further requires the model to recover both utterances in a single response and establish their correct spatial attribution. The original model achieves no recordings in which both speakers' content meets the success criterion. In contrast, BiSA Stage 2 reaches 58.85\% (678/1,152) both-speaker content success and 57.81\% (666/1,152) left-right attribution success. Overall WER decreases from 63.54\% to 14.02\%. The corresponding success rates for untrained binaural input are 5.12\% and 5.03\%. This finding addresses the gap between spatial perception and open-ended speech recovery. Effective semantic modeling and supervised adaptation of complementary binaural information help produce representations that support both content recovery and spatial attribution.

We further evaluate speech recovery under staggered overlap using BiCP-Async. In these recordings, one speaker begins speaking before the other joins. The model must recover the complete utterance of the spatially specified speaker while retaining the preceding speech of the first speaker in the input. Table~\ref{tab:bicp-async-results} reports transcription performance for the later-starting speaker, content confusion, and joint transcription results.

\begin{table*}[t]
\centering
\caption{Transcription performance under staggered overlap on BiCP-Async}
\label{tab:bicp-async-results}
\par\vspace{4pt}
\small
\setlength{\tabcolsep}{4pt}
\renewcommand{\arraystretch}{1.12}
\begin{tabularx}{\textwidth}{@{}l>{\centering\arraybackslash}X>{\centering\arraybackslash}X>{\centering\arraybackslash}X>{\centering\arraybackslash}X>{\centering\arraybackslash}X>{\centering\arraybackslash}X@{}}
\toprule
Model & Later-speaker WER $\downarrow$ & Earlier-speaker substitutions $\downarrow$ & Both-query success $\uparrow$ & Both-speaker content success $\uparrow$ & Left-right attribution success $\uparrow$ & Joint transcription WER $\downarrow$ \\
\midrule
\addlinespace[2pt]
Original model & 109.47 & 170 & 0.26 & 0.00 & 0.00 & 57.11 \\
Untrained binaural input & 104.08 & 89 & 5.73 & 5.73 & 5.47 & 55.82 \\
BiSA Stage 1 & 30.27 & \textbf{14} & 52.60 & 52.34 & 52.08 & \textbf{16.92} \\
BiSA Stage 2 & \textbf{29.58} & \textbf{14} & \textbf{53.39} & \textbf{53.65} & \textbf{53.65} & 18.03 \\
\bottomrule
\end{tabularx}
\par\smallskip
\begin{minipage}{\textwidth}
\footnotesize Note. The evaluation contains 384 complete staggered-overlap recordings. All values are percentages except earlier-speaker substitutions, which reports the number of recordings. An earlier-speaker substitution occurs when a response to the later-speaker transcription query has WER above 20\% against the later speaker's reference but at most 20\% against the earlier speaker's reference.
\end{minipage}
\end{table*}

For later-speaker transcription, the original model has a WER of 109.47\%, with 170 of 384 responses meeting the criterion for transcription of the earlier speaker instead. BiSA Stage 2 reduces later-speaker WER to 29.58\% and the number of earlier-speaker substitutions to 14/384. Both-query success on the same recording also increases from 0.26\% to 53.39\%. The sharp reduction in this error type reflects a more accurate correspondence between speech onset order and the target specified by spatial instructions. Even when the earlier speaker has established substantial speech context, the model can select the later-starting target according to spatial instructions and reduce interference from the preceding utterance during content recovery.

For joint transcription under staggered overlap, BiSA Stage 2 achieves 53.65\% (206/384) for both-speaker content success and left-right attribution success. Overall WER decreases from 57.11\% for the original model to 18.03\%. Together with the marked reduction in incorrect speaker selection during later-speaker transcription, the joint transcription performance reflects BiSA's improved alignment of speaker content with spatial positions in mixed speech. Spatial references guide target-utterance recovery more effectively, reducing the influence of speaking order on content selection and attribution.

\subsection{External Generalization and Preservation of Pretrained Capabilities}

The value of binaural semantic adaptation also lies in whether learned representations support target-utterance recovery across acoustic conditions and remain compatible with the pretrained model's audio understanding capabilities. We therefore evaluate BiSA from three perspectives: external speech recovery, general audio understanding, and spatial response consistency. We also examine how the two training stages affect these capabilities.

To assess cross-dataset applicability, we evaluate speech recovery on Clarity CEC1 \citep{graetzer2021clarity}. Originally developed for hearing-aid speech enhancement, CEC1 uses binaural room impulse responses to construct indoor acoustic scenes containing target speech and interfering sources. We select all 750 recordings with speech interference from the official evaluation set and use the mixtures from the front-facing microphones of the binaural hearing aids. Position-conditioned transcription prompts are constructed from the relative directions of the target and interfering speakers. We evaluate both full recordings and target-utterance windows. Full recordings retain the interfering speaker's speech before and after the target utterance. Target-utterance windows contain only the target utterance interval while preserving interference within that interval. WER is computed against the official target transcripts for all models. This evaluation introduces independent speech material and room acoustic conditions to assess BiSA in external speech interference scenarios.

\begin{table*}[t]
\centering
\caption{External speech recovery, pretrained capabilities, and spatial response consistency}
\label{tab:external-capabilities}
\par\vspace{4pt}
\small
\setlength{\tabcolsep}{4pt}
\renewcommand{\arraystretch}{1.12}
\begin{tabularx}{\textwidth}{@{}l>{\raggedright\arraybackslash}Xccc@{}}
\toprule
Evaluation dimension & Benchmark and metric & \shortstack{Original\\model} & \shortstack{BiSA\\Stage 1} & \shortstack{BiSA\\Stage 2} \\
\midrule
\addlinespace[2pt]
External speech recovery & CEC1 full-recording WER $\downarrow$ & 59.33 & \textbf{38.52} & 39.60 \\
 & CEC1 target-window WER $\downarrow$ & 27.01 & 19.35 & \textbf{18.66} \\
Spatial understanding & STAR-Bench spatial-reasoning subset accuracy $\uparrow$ & \textbf{70.25} & 68.33 & 69.06 \\
General audio understanding & MMAU-Pro overall score $\uparrow$ & \textbf{56.44} & 55.94 & 56.04 \\
Spatial understanding & MMAU-Pro spatial-question accuracy $\uparrow$ & 55.38 & 63.69 & \textbf{64.92} \\
Ear-swap consistency & STAR-Bench ear-swap pair accuracy $\uparrow$ & \textbf{72.56} & 66.17 & 70.30 \\
 & MMAU-Pro ear-swap pair accuracy $\uparrow$ & 33.11 & 31.08 & \textbf{33.78} \\
\bottomrule
\end{tabularx}
\par\smallskip
\begin{minipage}{\textwidth}
\footnotesize Note. WER and accuracy are reported as percentages; the MMAU-Pro overall score is reported on a 0--100 scale. CEC1 contains 750 official recordings with speech interference. Ear-swap evaluations contain 266 input pairs for STAR-Bench and 148 for MMAU-Pro. The MMAU-Pro evaluation protocol is detailed in Appendix~\ref{app:minimal-evaluation}. Ear-swap pair accuracy requires correct responses for both the original input and its channel-swapped counterpart.
\end{minipage}
\end{table*}

As shown in Table~\ref{tab:external-capabilities}, the original model obtains 59.33\% WER on full CEC1 recordings. BiSA reduces this to 38.52\% after Stage 1 and 39.60\% after Stage 2. On target-utterance windows, the corresponding WERs are 27.01\%, 19.35\%, and 18.66\%. The clear improvements under both conditions show that target-utterance recovery learned through binaural semantic modeling and supervised adaptation transfers to independent speech material and room acoustic conditions. In full recordings, BiSA also follows spatial instructions to recover target utterances more accurately despite the interfering speaker's preceding and trailing speech. This further demonstrates the effectiveness of the learned capability in longer contexts of competing speech.

We further use the 502-question spatial-reasoning subset of STAR-Bench \citep{liu2026star} and MMAU-Pro \citep{kumar2026mmau} to assess how binaural adaptation affects the pretrained model's audio understanding capabilities. BiSA Stage 2 achieves 69.06\% accuracy on the STAR-Bench spatial-reasoning subset, close to the original model's 70.25\%. On MMAU-Pro, spatial-question accuracy increases from 55.38\% to 64.92\%, while overall score reaches 56.04\%, close to the original model's 56.44\%. These results show that the capability expansion achieved through binaural semantic encoding, cross-ear interaction, and supervised adaptation improves spatial tasks while largely preserving general audio understanding performance. The effective use of binaural semantic information therefore offers benefits beyond multi-talker transcription.

To further examine the model's responses to changes in binaural spatial information, we evaluate paired response accuracy under physical ear swapping on STAR-Bench and MMAU-Pro. Swapping the left and right channels preserves speech content while changing the corresponding spatial direction relationships. From Stage 1 to Stage 2, ear-swap pair accuracy increases from 66.17\% to 70.30\% on STAR-Bench and from 31.08\% to 33.78\% on MMAU-Pro. Stage 2 freezes the shared low-rank adaptation parameters and cross-ear differential module, updating only the two ear-identity vectors. Further calibration of binaural input identities can therefore improve response consistency under changes in spatial input. Together with the external transcription and pretrained capability evaluations, the two training stages serve distinct adaptation roles. Stage 1 establishes binaural semantic recovery, while Stage 2 further adjusts the model's use of spatial information while largely retaining that capability.

\begin{table*}[t]
\centering
\caption{Ablation of cross-ear differential interaction on BiCP-Sync}
\label{tab:interaction-ablation}
\par\vspace{4pt}
\small
\setlength{\tabcolsep}{4pt}
\renewcommand{\arraystretch}{1.15}
\begin{tabularx}{\textwidth}{@{}l*{5}{>{\centering\arraybackslash}X}@{}}
\toprule
Method & \shortstack{Position-conditioned\\WER $\downarrow$} & \shortstack{Both-query\\success $\uparrow$} & \shortstack{Joint transcription\\WER $\downarrow$} & \shortstack{Both-speaker\\content success $\uparrow$} & \shortstack{Left-right\\attribution success $\uparrow$} \\
\midrule
\addlinespace[2pt]
No cross-ear interaction & 48.57 & 6.48 & 40.52 & 11.11 & 10.65 \\
Same-ear self-interaction & 44.33 & 11.11 & 37.64 & 14.35 & 13.43 \\
\textbf{Full BiSA Stage 1} & \textbf{17.04} & \textbf{57.41} & \textbf{13.81} & \textbf{58.80} & \textbf{58.33} \\
\bottomrule
\end{tabularx}
\par\smallskip
\begin{minipage}{\textwidth}
\footnotesize Note. All values are percentages. Evaluation uses a fixed, stratified subset of 216 simultaneous two-speaker recordings. The configuration without cross-ear interaction contains 1,973,248 trainable parameters. Same-ear self-interaction and full BiSA each contain 2,300,928 trainable parameters. All three configurations use the same training data, initialization protocol, and evaluation procedure. Results are based on a single training seed.
\end{minipage}
\end{table*}

\subsection{Ablation and Mechanism Analysis}
\label{sec:ablation}

To assess the contribution of binaural semantic interaction to multi-talker speech recovery, we construct two ablation baselines under the same training protocol. The first removes the cross-ear differential module, retaining shared binaural semantic encoding and low-rank adaptation. The second retains the full module structure but replaces cross-ear differencing with residual updates based on same-ear hidden states, controlling for additional parameter capacity. All configurations use the same training data and number of training steps. Evaluation is conducted on a fixed subset of 216 recordings from BiCP-Sync. Table~\ref{tab:interaction-ablation} reports the results for both transcription tasks.

As shown in Table~\ref{tab:interaction-ablation}, removing cross-ear interaction yields 6.48\% both-query success, with 11.11\% both-speaker content success and 10.65\% left-right attribution success in joint transcription. Same-ear self-interaction increases these three rates to 11.11\%, 14.35\%, and 13.43\%, respectively. In comparison, full BiSA achieves 57.41\% both-query success, 58.80\% both-speaker content success, and 58.33\% left-right attribution success. Its WER decreases to 17.04\% for position-conditioned transcription and 13.81\% for joint transcription. Under the same binaural inputs and supervision, explicitly modeling differences between the two ears' semantic representations supports more effective recovery and spatial attribution of both speakers' utterances than independent binaural encoding.

Compared with the parameter-matched same-ear self-interaction baseline, full BiSA improves both-query success by 46.30 percentage points. Both-speaker content success and left-right attribution success increase by 44.44 and 44.91 percentage points, respectively. The two configurations have identical parameter counts, module placement, and training conditions. The additional gains therefore reflect the contribution of cross-ear differential interaction beyond the effect of added parameter capacity. The value of binaural semantic information depends on both the complementary observations it provides and how their differences are used during encoding. Cross-ear differencing allows semantic representation updates to explicitly incorporate information from the opposite ear, providing an effective approach to content recovery and spatial attribution in overlapping speech.

\section{Conclusion}
\label{sec:conclusion}

We introduce BiCP-Bench and propose Binaural Semantic Adaptation (BiSA) to investigate how spatial audio models recover complete utterances from different speakers in overlapping speech. BiCP-Bench systematically characterizes the gap between spatial perception and complete utterance recovery through candidate-based utterance selection, position-conditioned transcription, and joint transcription in simultaneous and staggered overlap scenarios. Controlled reference-ear experiments further reveal the complementarity of binaural semantic information and the limitations of single-ear semantic encoding. Based on this finding, BiSA uses a shared semantic encoder for both ears and cross-ear differential interactions to combine complementary information during semantic representation construction.

Experiments on BiCP-Bench show that BiSA substantially improves complete utterance recovery and spatial attribution for multiple speakers under simultaneous and staggered overlap. These gains extend to independent speech material and acoustic conditions and remain evident in full recordings that retain the earlier-starting interfering speaker's speech. BiSA also largely preserves the pretrained model's general audio understanding performance while further improving performance on spatial tasks. These results reveal that reliably associating speaker positions with utterance content in overlapping speech depends on spatial perception and on how well semantic encoding preserves each speaker's speech content. By jointly modeling complementary binaural information, BiSA improves spatially grounded multi-talker speech understanding even with the original spatial encoding pathway unchanged.

\noindent\textbf{Code and Benchmark Availability.} The BiSA implementation, model weights, and BiCP-Bench are being prepared for public release.


\begingroup
\small\raggedright
\urlstyle{same}
\bibliography{references}
\endgroup

\clearpage
\appendix
\section{Task and Evaluation Details}
\label{app:minimal-evaluation}

\noindent\textbf{Formal prompts.} Table~\ref{tab:formal-prompts} gives the prompts used for the reported BiCP-Bench results. Braced fields denote the queried direction or candidate utterance. Candidate questions use absolute direction labels. Target transcription uses \emph{front-left} and \emph{front-right} for opposite-side BiCP-Sync mixtures, and relative \emph{left} and \emph{right} for same-side mixtures and BiCP-Async.

\begin{table}[!ht]
\centering
\caption{Formal BiCP-Bench task prompts.}
\label{tab:formal-prompts}
\small
\renewcommand{\arraystretch}{1.1}
\begin{tabularx}{\columnwidth}{@{}>{\raggedright\arraybackslash}X@{}}
\toprule
\textbf{Candidate-based content recognition} \\
Which statement is made by the speaker on the \{direction\} in this recording?\newline
(A) \{candidate\_A\}\newline
(B) \{candidate\_B\}\newline
(C) \{candidate\_C\}\newline
(D) \{candidate\_D\}\newline
Answer with the option letter only. \\
\midrule
\textbf{Position-conditioned target transcription} \\
Transcribe the speech of the speaker on the \{direction\}. Output only the words spoken by that speaker. \\
\midrule
\textbf{Joint transcription with spatial attribution} \\
Transcribe both speakers verbatim. Use their relative positions: the more leftward speaker is left, and the more rightward speaker is right. Output exactly two lines in the format:\newline
left: ...\newline
right: ... \\
\bottomrule
\end{tabularx}
\end{table}

\newpage
\noindent\textbf{Scoring details.} WER scoring lowercases text and ignores punctuation, retaining word-internal apostrophes. Target transcription scores the entire generated response.

The content-coverage diagnostics in Sections~\ref{sec:reference-ear} and~\ref{sec:ear-complementarity} ignore spatial labels and allow extra output segments. Formal BiCP-Bench content success requires no unmatched output words and evaluates explicit left--right attribution separately, as defined in Section~\ref{sec:bicp-evaluation}.

\noindent\textbf{External evaluation.} The reported STAR-Bench accuracy covers its 502-question spatial-reasoning subset, averaged over three option orders per question. The MMAU-Pro overall score combines multiple-choice accuracy, instruction-following scores, and normalized open-ended judge scores, weighted by the number of evaluated questions.

BiSA keeps the pretrained audio projector and language model frozen, with added adaptation parameters saved and loaded separately. This permits a choice between the original and adapted configurations. In MMAU-Pro evaluation, we select saved responses and scores according to channel equality. Questions with identical left and right waveforms use the original model's saved response and score; other questions use BiSA's. We aggregate these selected results using the official scoring protocol.

\end{document}